\documentclass[preprint,12pt]{elsarticle}

\usepackage{amssymb}
\usepackage{amsmath}
\journal{Ultrasound in Medicine \& Biology}

\usepackage{xcolor}
\usepackage{hyperref}
\hypersetup{colorlinks=true}
\usepackage{booktabs}
\usepackage{multirow}
\usepackage[table,xcdraw]{xcolor}
\usepackage{ulem}
\usepackage{adjustbox}
\usepackage{bm}
\usepackage{threeparttable}
\usepackage{lineno}
\useunder{\uline}{\ul}{}
\begin{document}

\setcounter{secnumdepth}{0}

% Exclude front matter from word count
%TC:ignore 
\begin{frontmatter}



\title{MonoUNet: A Robust Tiny Neural Network for Automated Knee Cartilage Segmentation on Point-of-Care Ultrasound Devices}


\author[1]{Alvin Kimbowa}
\author[2]{Arjun Parmar}
\author[3]{Ibrahim Mujtaba}
\author[3]{Will Wei}
\author[4]{Maziar Badii}
\author[2]{Matthew Harkey}
\author[3]{David Liu}
\author[3,5]{Ilker Hacihaliloglu}

\affiliation[1]{organization={School of Biomedical Engineering, The University of British Columbia},%Department and Organization
            % city={},
            country={Canada}}
\affiliation[2]{organization={Department of Kinesiology, Michigan State University},%Department and Organization
            % city={},
            country={USA}}
\affiliation[3]{organization={Department of Rheumatology, The University of British Columbia},%Department and Organization
            % city={},
            country={Canada}}
\affiliation[4]{organization={Department of Radiology, The University of British Columbia},%Department and Organization
            % city={},
            country={Canada}}
\affiliation[5]{organization={Department of Medicine, The University of British Columbia},%Department and Organization
            % city={},
            country={Canada}}

%% Abstract
\begin{abstract}
\textbf{Objective:}
To develop a robust and compact deep learning model for automated knee cartilage segmentation on point-of-care ultrasound (POCUS) devices.

\textbf{Methods:}
We propose MonoUNet, a novel, highly compact segmentation model consisting of (i) an aggressively reduced U-Net backbone, (ii) a trainable monogenic block that extracts multi-scale local phase features from the input, and (iii) a gating mechanism that injects these features into the encoder stages to reduce sensitivity to variations in ultrasound image appearance.
MonoUNet segmentation performance was evaluated on a multi-site, multi-device knee cartilage ultrasound dataset using Dice score and mean average surface distance (MASD).
Agreement between MonoUNet and manual cartilage outcomes (thickness and echo intensity) was assessed using Bland–Altman analysis with $95\%$ limits of agreement, and reliability was assessed using intraclass correlation coefficient (ICC$_{2,k}$).

\textbf{Results:}
Overall, MonoUNet outperformed existing lightweight segmentation models, with average Dice scores ranging from 92.62\% to 94.82\% and MASD values between 0.133 mm and 0.254 mm.
MonoUNet reduces the number of parameters by 10x--700x and computational cost by 14x--2000x relative to existing lightweight models.
MonoUNet cartilage outcomes showed excellent reliability and agreement with the manual outcomes: intraclass correlation coefficients (ICC\(_{2,k}\))=0.96 and bias=2.00\% (0.047 mm) for average thickness, and ICC\(_{2,k}\)=0.99 and bias=0.80\% (0.328 a.u.) for echo intensity.

\textbf{Conclusion:}
Incorporating trainable local phase features improves the robustness of highly compact neural networks for knee cartilage segmentation across varying acquisition settings and could support scalable ultrasound-based assessment and monitoring of knee osteoarthritis using POCUS devices.
\textit{The code is publicly available at \href{https://github.com/alvinkimbowa/monounet}{GitHub}}.

\end{abstract}

% \includegraphics[width=0.75\linewidth]{graphical_abstract.pdf}

% \begin{highlights}
% \item A lightweight neural network is proposed for automated knee cartilage segmentation in ultrasound.
% \item Trainable multi-scale local phase features are integrated into a U-Net–based architecture.
% \item The proposed model achieves robust cross-device performance on multi-site ultrasound data.
% \item Accurate cartilage thickness and intensity measurements are obtained from automated segmentations.
% \end{highlights}

%% Keywords

\end{frontmatter}
%TC:endignore



\section{Introduction}
\label{sec:introduction}
% What is the problem?
Knee osteoarthritis (OA) is the most common joint disease affecting over 650 million people globally, and is a leading cause of disability in adults~\cite{cui_global_2020}.
Despite its prevalence, knee OA has no cure, is often diagnosed at a late stage, and its pathophysiology remains unclear~\cite{he_pathogenesis_2020}.
Knee OA is characterized by progressive degeneration and thinning of the cartilage, and monitoring these morphological changes may enable early detection in high-risk groups such as individuals with prior knee injuries, and older adults~\cite{emery_establishing_2019,collins_magnetic_2025}.

Magnetic Resonance Imaging (MRI) is currently the gold standard for early imaging of knee OA~\cite{oei_osteoarthritis_2022}.
However, its high cost and long wait times make it impractical for routine knee OA screening, frequent monitoring, and large-scale or longitudinal studies in community-based settings~\cite{roemer_imaging_2022}.
In contrast, point-of-care ultrasound (POCUS) offers a low-cost, non-invasive, portable, and widely accessible imaging modality capable of visualizing knee joint structures, including the cartilage~\cite{roemer_imaging_2022}.
As a result, POCUS is increasingly being adopted in both clinical practice and knee OA research for cartilage assessment, disease progression analysis, and treatment response evaluation~\cite{nakashima_point--care_2022,harkey_femoral_2024,dagostino_ultrasound_2024,parmar_wireless_2024}.
% These properties make ultrasound particularly attractive for population-level studies and repeated measurements over time.

Accurate and robust cartilage segmentation is a critical prerequisite for both clinical and research-based ultrasound analysis of knee OA. 
However, ultrasound-based segmentation remains challenging due to a combination of anatomical, physical, and acquisition-related factors.
Unlike many imaging modalities where tissue boundaries are well defined, cartilage interfaces in ultrasound images are often poorly delineated and smooth rather than forming sharp, high-contrast edges~\cite{desai_knee-cartilage_2019,zhang_research_2024}.
This intrinsic ambiguity complicates both manual annotation and automated segmentation.

Furthermore, ultrasound acquisition is highly operator-dependent~\cite{dagostino_ultrasound_2024,harkey2022validating}.
Even minor variations in probe handling, such as subtle tilting or changes in insonation angle, can lead to noticeable alterations in the visualized anatomy, including partial loss, distortion, or apparent displacement of cartilage boundaries.
Differences in acquisition parameters, such as frequency, focus depth, gain, and imaging presets, introduce additional heterogeneity.
These variations directly impact manual segmentation consistency and introduce substantial inter- and intra-rater variability in the measurements~\cite{papernick_reliability_2020}.
These challenges are amplified in real-world POCUS workflows, where imaging is performed by operators with varying levels of expertise and under diverse acquisition conditions.

The problem is further compounded by device-specific differences in image formation pipelines.
Ultrasound systems apply proprietary and often undocumented internal processing steps, including beamforming strategies, dynamic range compression, speckle characteristics, and post-processing filters.
As a result, the appearance and quality of cartilage structures vary significantly across devices~\cite{salimi_ultrasound_2022,perez-sanchez_comparison_2024}.
Ultrasound systems also differ substantially in hardware design, ranging from cart-based systems with high image quality, to portable laptop-based systems that balance performance and mobility, and handheld devices that prioritize accessibility but often exhibit lower image quality and increased noise~\cite{noauthor_esr_2019,rykkje_hand-held_2019}.

In both clinical and research settings, practical deployment constraints further motivate compact and efficient segmentation models.
Many small clinics, community health centers, and research sites lack access to high-performance computing infrastructure or reliable cloud-based processing due to cost, connectivity, latency, and data governance constraints~\cite{ahmed_integrating_2025}.
In addition, research studies often require portable ultrasound systems paired with tablets or lightweight workstations for data collection in community, outpatient, or field-based environments~\cite{ahmed_integrating_2025}.
In such scenarios, segmentation algorithms must operate locally and efficiently without dependence on external compute resources.

Real-time or near–real-time inference is therefore essential across both domains. In clinical workflows, immediate feedback during image acquisition can support quality assurance and point-of-care decision-making.
In research settings, real-time segmentation can enable standardized data collection, reduce post-processing burden, and improve consistency across operators and sites in large or longitudinal studies.
Achieving these goals requires compact neural network architectures that are robust to ultrasound variability while remaining computationally efficient and suitable for deployment on resource-constrained edge devices.

Various deep learning-based methods have been proposed to automate knee cartilage segmentation in ultrasound~\cite{desai_knee-cartilage_2019,desai_international_2021,antico_deep_2020}.
However, they are computationally expensive and do not align with POCUS deployment constraints, where on-device inference, limited computation, memory, and power efficiency are essential.
Recent work in the broader medical imaging literature has proposed lightweight neural network architectures with reduced parameter counts for efficient on-device performance~\cite{valanarasu_unext_2022,kalkhof_med-nca_2023,chen_tinyu-net_2024}.
However, compact architectures inherently have limited representational capacity, restricting their ability to learn robust, invariant features directly from B-mode ultrasound intensity data.
As a result, they often struggle to generalize to images from varying acquisition settings, limiting their clinical utility.

To address this limitation, efficient contrast and intensity invariant feature representations are required.
Local phase features naturally satisfy these properties~\cite{hacihaliloglu_2a-4_2006}.
Unlike intensity-based representations, local phase features capture structural information that is less sensitive to gain, attenuation, speckle statistics, and vendor-specific post-processing~\cite{hacihaliloglu_bone_2009}.
These properties make the features particularly robust to the ultrasound-specific challenges outlined above, and have been shown to improve ultrasound image analysis, including knee cartilage segmentation~\cite{desai_knee-cartilage_2019,harkey2022validating}.

In this paper, we propose MonoUNet, a novel, highly compact U-Net–based architecture that incorporates trainable multi-scale local phase features for robust knee cartilage segmentation on POCUS devices.
MonoUNet injects the local phase features into the high-resolution stages of the encoder, via a gating mechanism, to modulate the encoder features towards robust structural information.
By explicitly embedding local phase information into the network, the proposed design compensates for the limited feature extraction capacity of compact models while improving robustness to the dominant sources of variability in POCUS.
MonoUNet was extensively evaluated on a multi-site, multi-device knee cartilage ultrasound dataset to assess its performance.

\section{Materials and Methods}

\begin{figure}[]
    \centering
    \includegraphics[width=\linewidth]{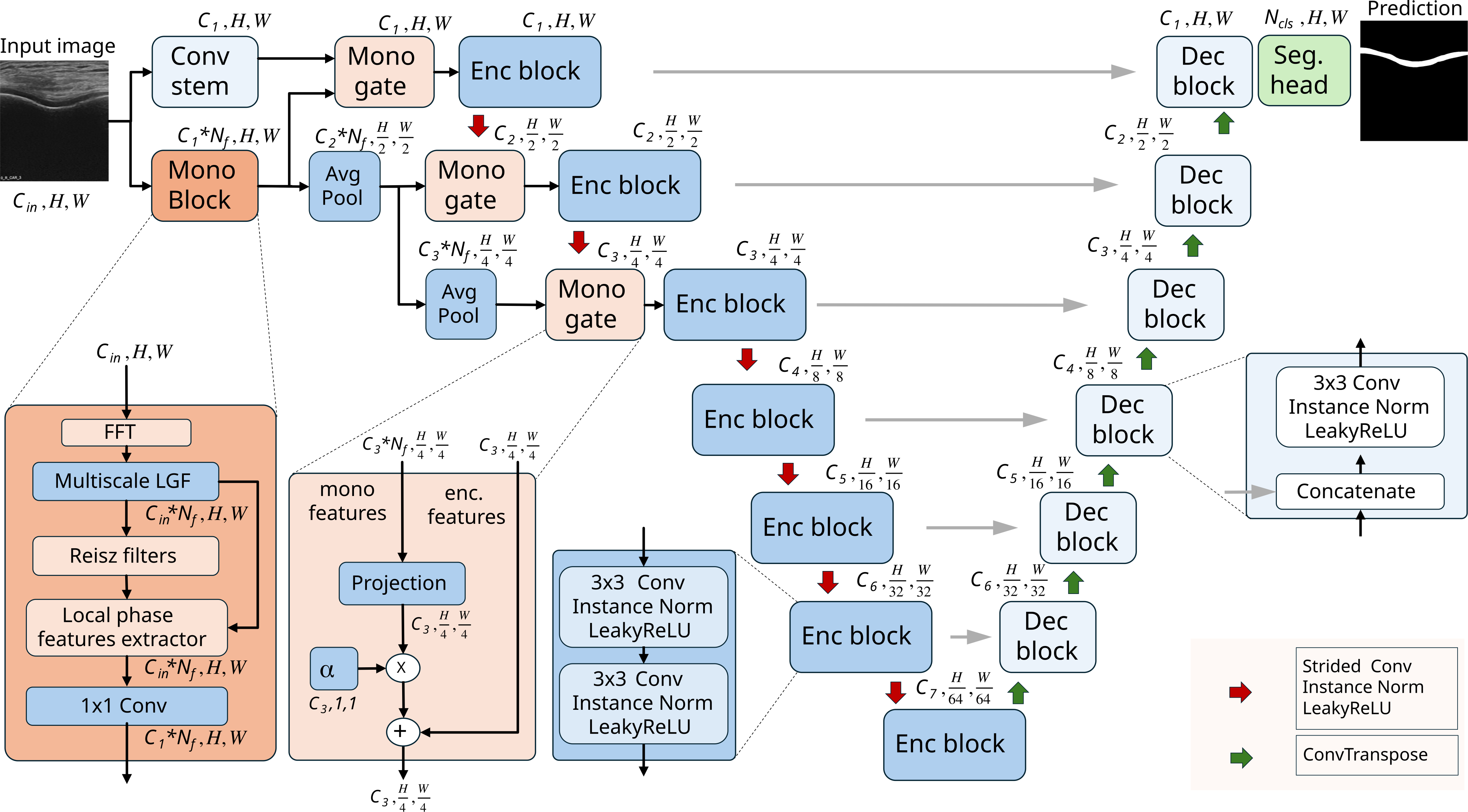}
    \caption{\textbf{Overview of MonoUNet architecture}: Trainable multi-scale local phase features are extracted from the input image using the Mono block and injected into the high-resolution encoder stages via Mono gates, where they are fused with the encoder features using learnable channel-wise weights. The decoder is half the size of the encoder.}
    \label{fig:arch}
\end{figure}

Fig.~\ref{fig:arch} shows the overall MonoUNet architecture.
MonoUNet builds upon a U-Net backbone architecture obtained from the self-configuring nnU-Net~\cite{isensee_nnu-net_2021} framework with modifications aimed at making the model compact.
This includes reducing the number of parameters in the base architecture, incorporating trainable local phase feature extractors using the Mono block, and injecting the features into the high-resolution stages of the U-Net encoder via the Mono gate.
We detail the different architectural components in the following subsections.

\subsection{MonoUNetBase}
We obtained the base U-Net configuration using the self-configuring nnU-Net framework~\cite{isensee_nnu-net_2021}.
The base U-Net has approximately 46 million parameters.
However, for real-time or pseudo-real-time on-device inference, we determined the desired model size to be below \(3{,}500\) parameters by empirically testing models of varying sizes within the image processing pipeline of a POCUS device.
Therefore, to reduce the base U-Net model size, we first asymmetrically reduced the decoder to a single convolutional block per stage (instead of two blocks as in the encoder) inspired by the residual nnU-Net configuration~\cite{isensee_nnu-net_2024}.
We then further aggressively reduced the parameter count by halving the number of channels following~\citet{hassler_lean_2025} to yield a model with a constant number of feature channels, \(C=2\), across all stages (i.e., \(C=C_1=C_2=C_3=C_4=C_5=C_6=C_7=2\)).
We refer to this model configuration as \textit{MonoUNetBase} and it has about \(1{,}140\) parameters.
For comparison, a configuration with C=4 yields a model with \(4{,}300\) parameters, which is above the desired model size.

\subsection{Mono block}
The Mono block is where local phase features are extracted from the input image using a trainable monogenic layer composed of multiple log-Gabor bandpass filters (LGFs)~\cite{moya-sanchez_trainable_2021}.
Each LGF is parameterized by a learnable center frequency $\omega_0$, bandwidth parameter $\sigma_r$, and a geometric scaling factor $r$, and produces $M$ multi-scale responses.
Eq.~\ref{eqn:log-gabor} defines the frequency-domain LGF response at a given scale.
\begin{align}
    LGF(\bm{\omega}, \omega_{0,m},\sigma_r) &= \exp \left(
    - \frac{
    \log^2\left(\frac{|\bm{\omega}|}{\omega_{0,m}}\right)
    }
    {
    2\log^2\left(\sigma_r\right)
    }
    \right),
    \label{eqn:log-gabor}
\end{align}
where $\bm{\omega} = (\omega_x, \omega_y)$ denotes the 2D frequency vector, $\omega_{0,m} = \omega_0 r^{-(m-1)}$ is the center frequency of the $m^{th}$ scale filter, $m=1,2,...,M$, and $r > 1$ is the learned scaling factor.
The Mono block learns $k$ such LGFs resulting in a total of \(N_f=k\times M\) features for each input channel.
To minimize the number of added parameters, we set \(M=3\) guided by empirical results showing no significant improvement with additional scales and is consistent with best practice~\cite{hacihaliloglu_automatic_2011}.
We set k equal to the number of encoder stages in which the local phase features are to be injected, essentially learning a single LGF per stage.
Empirically, injecting features in deeper layers yields diminishing returns, as these layers have largely lost structural information.
We, therefore, set $k=3$, corresponding to the high-resolution encoder stages.

Riesz filters, $R_1$ and $R_2$, are then applied to the LGF-filtered images, $I_e$, to obtain the quadrature components of the monogenic signal, from which the local phase features, $I_\theta$, are extracted following Eq.~\ref{eqn:local_phase}.
\begin{align}
    I_\theta(x,y) &= \arctan\!\left(
    \frac{I_e}{\sqrt{I_{o1}^2 + I_{o2}^2}}
    \right), \label{eqn:local_phase}
\end{align}
where
\begin{align*}
I_e(x,y) &= \mathcal{F}^{-1}\!\left\{
LGF \odot \mathcal{F}\{I\}
\right\}, \\
I_{o1}(x,y) &= \mathcal{F}^{-1}\!\left\{
R_1 \odot \mathcal{F}\{I_e\}
\right\}, \\
I_{o2}(x,y) &= \mathcal{F}^{-1}\!\left\{
R_2 \odot \mathcal{F}\{I_e\}
\right\}, \\
R_1(\omega_x, \omega_y) &= i\frac{\omega_x}{\sqrt{\omega_x^2 + \omega_y^2}},\\
R_2(\omega_x, \omega_y) &= i\frac{\omega_y}{\sqrt{\omega_x^2 + \omega_y^2}},
\end{align*}
where $I$ is the input image, $\mathcal{F}$ is the Fourier Transform, $\odot$ is element-wise multiplication, and  $\omega_x$ and $\omega_y$ denote the horizontal and vertical frequency components, respectively.
The multi-scale local phase features are combined via a pointwise (1×1) convolution to allow the network to learn scale-dependent weighting of the phase responses to extract richer representations.

\subsection{Mono gate}
The local phase features are injected into the high-resolution encoder stages via the Mono gate.
Except for the first encoder stage, the phase features are first downsampled using average pooling to match the spatial resolution of the corresponding encoder stage.
Inside the Mono gate, a pointwise (1x1) convolution is used to project the phase features to the same channel dimension as the corresponding encoder features.
The phase and encoder features are combined using a channel-wise weighted sum, allowing the network to adaptively control the contribution of the local phase features via learnable weights $\alpha$.

\subsection{Dataset overview}
We used four privately collected 2D ultrasound knee cartilage datasets including three retrospective datasets (D0, D1, D2) and one prospective dataset (D3), as summarized in Table~\ref{tab:dataset}.

The retrospective datasets were acquired from 35 subjects who had undergone anterior cruciate ligament reconstruction (ACLR), and 192 healthy volunteers at sites A and B, using three ultrasound devices.
D0 was acquired using a cart-based system (GE LOGIQ P9 R3 ultrasound system with the L3-12-RS wideband linear array probe), D1 using a portable laptop-based system (GE LOGIQ e ultrasound system with a 12 MHz linear probe), and D2 using a portable handheld ultrasound system (Clarius HD3 L15).
Some healthy subjects (n=71) were imaged with both the GE LOGIQ P9 R3 and the Clarius HD3 L15 at the same visit.
Subjects were positioned supine with the knee fully flexed, and the transducer was placed transversely with the femoral intercondylar notch centered.
Three images were obtained per knee at a fixed imaging depth of 4 cm.
To assess inter-rater agreement, two independent annotators were recruited and trained to re-annotate a randomly selected subset of the retrospective images (n = 73), blinded to the original annotations.

% \usepackage{booktabs}
% \usepackage{multirow}
\begin{table}[]
\centering
\caption{\textbf{Dataset overview:} The retrospective datasets (D0-D2) were collected with three ultrasound devices, and the prospective dataset (D3) collected with one ultrasound device.}
\label{tab:dataset}
\begin{adjustbox}{max width=\textwidth}
\begin{threeparttable}
\centering
\begin{tabular}{@{}ccclccccc@{}}
\toprule
\multicolumn{1}{l}{\multirow{2}{*}{\textbf{Dataset}}} & \multicolumn{1}{l}{\multirow{2}{*}{\textbf{Site}}} & \multicolumn{1}{l}{\multirow{2}{*}{\textbf{Device type}}} & \multirow{2}{*}{\textbf{Devices}} & \multicolumn{3}{c}{\textbf{\# Subjects}} & \multicolumn{2}{c}{\textbf{\# Images}} \\ \cmidrule(lr){5-7} \cmidrule(l){8-9}
\multicolumn{1}{l}{} & \multicolumn{1}{l}{} & \multicolumn{1}{l}{} &  & \textbf{ACL} & \textbf{Healthy} & \textbf{Total} & \textbf{Total} & \textbf{Inter-rater} \\ \midrule
D0 & A & CB\tnote{a} & GE LOGIQ P9 R3 & 15 & 136 & 151 & 1787 & 26 \\
D1 & B & PL\tnote{b} & GE LOGIQ e & 20 & 56 & 76 & 587 & 22 \\
D2 & A & PH\tnote{c} & Clarius HD3 L15 & 0 & 71\tnote{*} & 71\tnote{*} & 234 & 25 \\
D3 & C & PH\tnote{c} & Viatom Dual Head & 0 & 1 & 1 & 400 & - \\ \midrule
\textbf{Total} & \textbf{-} & \textbf{-} & \multicolumn{1}{c}{\textbf{-}} & \textbf{35} & \textbf{193} & \textbf{228} & \textbf{3008} & \textbf{73} \\ \bottomrule
\end{tabular}
\begin{tablenotes}[flushleft]
\footnotesize
\item[a] CB = Cart-based ultrasound system.
\item[b] PL = Portable laptop-based ultrasound system.
\item[c] PH = Portable handheld ultrasound system.
\item[*] D2 subjects are a subset of D0 subjects, i.e., they were imaged with both the LOGIQ P9 R3 and the Clarius HD3 L15 at the same visit.
\end{tablenotes}
\end{threeparttable}
\end{adjustbox}
\end{table}

The prospective dataset D3 was collected at site C to further assess the generalizability of MonoUNet under varied acquisition conditions.
Four ultrasound videos (100 frames each) were acquired from a single healthy volunteer using a portable handheld system (Viatom Dual Head Scanner).
Data acquisition followed a protocol similar to that of the retrospective dataset, with additional variations in probe orientation, including intentional probe tilting to simulate sub-optimal acquisition conditions.
% Individual ultrasound images were extracted from the videos resulting in a total of 400 images.

\subsection{Experiments}
We consider three practical training scenarios that reflect common clinical settings for knee ultrasound: 1) access to a large, high-quality labeled dataset acquired with a high-end ultrasound device (D0), 2) access to a smaller, medium-quality dataset collected with a portable laptop-based ultrasound system (D1), and 3) access to only a limited, low-quality dataset acquired with a handheld POCUS device (D2).
In each scenario, the goal is to train a model that generalizes to the remaining datasets.

For each scenario, the data was randomly split into train and validation sets using an 80/20 split.
To minimize bias due to the random split and model initialization, we repeated the splitting and training process three times using different random seeds, resulting in three independently trained models per training dataset.
Each model was then evaluated on the remaining datasets to assess generalizability.

\subsection{Training details}
We resized the images to 256x256 followed by z-score normalization.
To encourage the Mono block to learn multi-scale features, we used affine rotation (\(\pm15^{\circ}\)) and scaling (0.8 - 1.2) with a probability of 0.8.
We trained MonoUNet for 1000 epochs using the AdamW optimizer~\cite{loshchilov_decoupled_2019} with a weight decay of 0.01, an initial learning rate of 0.01, a polynomial learning rate scheduler with a power of 0.9, a batch size of 8, and a combined binary cross entropy and Dice loss~\cite{azad_loss_2023}.
We chose the model with the best Dice on the validation set.
We also trained existing state-of-the-art lightweight deep learning models including UNeXt~\cite{valanarasu_unext_2022}, CMUNeXt~\cite{tang_cmunext_2023}, Med-NCA~\cite{kalkhof_med-nca_2023}, and TinyU-Net~\cite{chen_tinyu-net_2024} as baselines for comparison with MonoUNet.
The models were trained on similar dataset splits using their open-source implementations and training protocols described in the original publications.
All experiments were conducted in PyTorch 2.8~\cite{paszke_pytorch_2019} on a single NVIDIA Tesla V100 (32 GB VRAM) graphics processing unit (GPU).

\subsection{Evaluation metrics}

We evaluated MonoUNet performance against manual labels using the Dice similarity coefficient as a measure of spatial overlap, computed according to Eq.~\ref{eqn:dice}.
We also used the mean average surface distance (MASD) to quantify boundary agreement, following~\citet{maier-hein_metrics_2024} and computed using Eq.~\ref{eqn:masd}.

\begin{equation}
\text{Dice} = \frac{2\,\text{TP}}{2\,\text{TP} + \text{FP} + \text{FN}},
\label{eqn:dice}
\end{equation}
where TP denotes true positives, FP false positives, and FN false negatives.

\begin{equation}
\text{MASD}(A, B) =
\frac{1}{2}
\left(
\frac{1}{|A|}
\sum_{a \in A} d(a, B)
+
\frac{1}{|B|}
\sum_{b \in B} d(b, A)
\right),
\label{eqn:masd}
\end{equation}
where \(A\) and \(B\) denote the boundaries extracted from the automated and manual segmentation masks, respectively, \(d(a, B) = \min_{b \in B} \lVert a - b \rVert_2\) is the Euclidean distance from a point \(a\) on boundary \(A\) to the closest point on boundary \(B\), and \(d(b, A) = \min_{a \in A} \lVert b - a \rVert_2\) is the Euclidean distance from a point \(b\) on boundary \(B\) to the closest point on boundary \(A\).
Note that we consider the largest connected component as the final prediction for all models.
Empty predictions were excluded when computing MASD as they yield undefined values.
Model efficiency was assessed using number of parameters and computational cost (floating point operations (FLOPs)).

\subsection{Statistical Analysis}
To assess the agreement between MonoUNet and manual outcomes, we computed the average cartilage thickness and echo intensity following~\citep{harkey2022validating}.
% These included average cartilage thickness and average cartilage intensity for the medial, lateral and intercondylar cartilage.
% The cartilage was sectioned into the three regions similar to the approach in~\cite{lisee_reliability_2020}, except that we automatically selected the mid point of the cartilage as the average of the lowest points in the cartilage.
% We performed this experiment on a subset of dataset D0 similar to the data used in~\cite{harkey2022validating} for comparison purposes.
We used Bland-Altman plots, with 95\% limits of agreement (LOA), to assess agreement, and the two-way random effects intraclass correlation coefficients based on absolute agreement ($ICC_{2,k}$) with $95\%$ confidence intervals (CI) to evaluate reliability.
ICC values less than 0.5 were considered poor reliability, values between 0.75 and 0.9 were considered good reliability, and values greater than 0.90 were considered excellent reliability~\cite{harkey2022validating}.
We used the model trained on the first split of dataset D1 and evaluated on D2, reflecting a realistic POCUS deployment scenario in knee ultrasound imaging.

% \subsection{Ablation studies}
% To evaluate the contribution of each proposed module, we conducted ablation experiments by incrementally adding one module at a time to the baseline model.
% We trained the models on dataset D1 and evaluated them on dataset D2 to assess cross-device generalizability, following the same setup in the \textit{Training Details} subsection.

\section{Results}

Tables~\ref{tab:train_d0},~\ref{tab:train_d1}, and~\ref{tab:train_d2}, and Figs.~\ref{fig:dice_distribution} and~\ref{fig:qualitative} summarize the average cross-dataset segmentation performance of MonoUNet and baseline models trained on D0, D1, and D2, respectively.
The tables also include inter-rater agreement against the retrospective manual annotations.

\subsection{Quantitative results}
MonoUNet had the smallest model size, with \(1,390\) parameters, and computational cost of 0.15G floating point operations (FLOPs).
This is 12.5x fewer parameters compared to the next smallest model, Med-NCA (70 thousand parameters), and 14.5x more computationally efficient compared to the next efficient model, CMUNeXt-S (2.18 GFLOPs).

% \usepackage{booktabs}
% \usepackage{multirow}
% \usepackage[normalem]{ulem}
\begin{table}[]
\centering
\caption{Cross-dataset segmentation performance for models trained on D0 and evaluated on D1, D2, and D3. Results are reported as mean ± standard deviation for Dice and MASD (mm). Best results for each column are highlighted in bold and second-best are underlined. Inter-rater variability is measured against the retrospective manual segmentations.}
\label{tab:train_d0} 
\begin{adjustbox}{max width=\textwidth}
\begin{tabular}{@{}lcccccccc@{}}
\toprule
\multirow{2}{*}{\textbf{Config.}} & \multirow{2}{*}{\textbf{\begin{tabular}[c]{@{}c@{}}Params\\ (k)$\downarrow$\end{tabular}}} & \multirow{2}{*}{\textbf{\begin{tabular}[c]{@{}c@{}}FLOPS\\ (G)$\downarrow$\end{tabular}}} & \multicolumn{2}{c}{\textbf{D1}} & \multicolumn{2}{c}{\textbf{D2}} & \multicolumn{2}{c}{\textbf{D3}} \\ \cmidrule(l){4-9} 
 &  &  & \textbf{Dice}$\uparrow$ & \textbf{MASD}$\downarrow$ & \textbf{Dice}$\uparrow$ & \textbf{MASD}$\downarrow$ & \textbf{Dice}$\uparrow$ & \textbf{MASD}$\downarrow$ \\ \midrule
UNeXt~\cite{valanarasu_unext_2022} & 1471.94 & 4.59 & 58.13±22.79 & 0.536±0.949 & 91.00±9.57 & {\ul 0.127±0.061} & 32.72±13.94 & 0.822±1.125 \\
Med-NCA~\cite{kalkhof_med-nca_2023} & {\ul 70.02} & 310.31 & 42.67±24.69 & 1.555±3.170 & 91.70±9.26 & \textbf{0.122±0.036} & {\ul 84.58±12.47} & {\ul 0.234±0.274} \\
CMUNeXt-S~\cite{tang_cmunext_2023} & 417.50 & {\ul 2.18} & 80.71±12.83 & 0.303±0.577 & {\ul 93.17±6.11} & 0.138±0.040 & 73.70±15.36 & 0.333±0.290 \\
TinyU-Net~\cite{chen_tinyu-net_2024} & 481.17 & 3.33 & {\ul 90.30±5.96} & {\ul 0.180±0.083} & 92.25±7.08 & 0.151±0.047 & 57.85±16.18 & 0.448±0.448 \\
\textbf{MonoUNet} & \textbf{1.39} & \textbf{0.15} & \textbf{93.02±4.08} & \textbf{0.160±0.094} & \textbf{93.77±4.98} & 0.146±0.070 & \textbf{94.82±2.35} & \textbf{0.202±0.204} \\ \midrule
Rater 1 & - & - & 93.67±1.92 & 0.133±0.034 & 93.21±2.45 & 0.139±0.048 & - & - \\
Rater 2 & - & - & 91.37±3.42 & 0.192±0.067 & 92.35±2.50 & 0.169±0.055 & - & - \\
\bottomrule
\end{tabular}
\end{adjustbox}
\end{table}

When trained on D0 (Table~\ref{tab:train_d0}), MonoUNet achieved Dice scores between 93.02\% and 94.82\% with MASD values below 0.21 mm across all test datasets, comparable to inter-rater variability on D1 and D2.
MonoUNet outperformed all baselines across all metrics, except for MASD on D2, where CMUNeXt-S and Med-NCA achieved marginally lower values (\(0.127\pm0.061\) mm and \(0.122\pm0.036\) mm, respectively, vs. \(0.146\pm0.070\) mm).
While some baselines, such as TinyU-Net and CMUNeXt-S, achieved competitive performance on individual datasets (such as, D1 and D2), their performance degraded substantially under stronger domain shifts, particularly on D3 where TinyU-Net and CMUNeXt-S achieved Dice scores of \(57.89\pm16.24\%\) and \(73.70\pm15.36\%\), respectively.

% \usepackage{booktabs}
% \usepackage{multirow}
% \usepackage[normalem]{ulem}
\begin{table}[]
\centering
\caption{Cross-dataset segmentation performance for models trained on D1 and evaluated on D0, D2, and D3. Results are reported as mean ± standard deviation for Dice and MASD (mm). Best results for each column are highlighted in bold and second-best are underlined. Inter-rater variability is measured against the retrospective manual segmentations.}
\label{tab:train_d1} 
\begin{adjustbox}{max width=\textwidth}
\begin{tabular}{@{}lcccccccc@{}}
\toprule
\multirow{2}{*}{\textbf{Config.}} & \multirow{2}{*}{\textbf{\begin{tabular}[c]{@{}c@{}}Params\\ (k)$\downarrow$\end{tabular}}} & \multirow{2}{*}{\textbf{\begin{tabular}[c]{@{}c@{}}FLOPS\\ (G)$\downarrow$\end{tabular}}} & \multicolumn{2}{c}{\textbf{D0}} & \multicolumn{2}{c}{\textbf{D2}} & \multicolumn{2}{c}{\textbf{D3}} \\ \cmidrule(l){4-9} 
 &  &  & \textbf{Dice}$\uparrow$ & \textbf{MASD}$\downarrow$ & \textbf{Dice}$\uparrow$ & \textbf{MASD}$\downarrow$ & \textbf{Dice}$\uparrow$ & \textbf{MASD}$\downarrow$ \\ \midrule
UNeXt~\cite{valanarasu_unext_2022} & 1471.94 & 4.59 & 68.10±20.30 & 0.208±0.298 & 38.54±24.41 & 0.934±1.837 & 5.20±8.75 & 2.816±1.964 \\
Med-NCA~\cite{kalkhof_med-nca_2023} & {\ul 70.02} & 310.31 & 88.23±8.84 & 0.152±0.149 & 71.92±14.09 & 0.304±0.303 & 25.28±5.12 & 2.113±0.502 \\
CMUNeXt-S~\cite{tang_cmunext_2023} & 417.50 & {\ul 2.18} & 92.27±5.32 & 0.148±0.050 & 73.33±22.41 & 0.367±0.666 & 16.13±13.38 & 1.741±1.220 \\
TinyU-Net~\cite{chen_tinyu-net_2024} & 481.17 & 3.33 & {\ul 93.57±2.61} & {\ul 0.147±0.046} & {\ul 87.88±12.26} & {\ul 0.190±0.190} & {\ul 39.83±16.87} & {\ul 1.218±1.244} \\
\textbf{MonoUNet} & \textbf{1.39} & \textbf{0.15} & \textbf{94.68±1.61} & \textbf{0.133±0.040} & \textbf{94.14±2.05} & \textbf{0.149±0.038} & \textbf{93.09±2.51} & \textbf{0.254±0.201} \\ \midrule
Rater 1 & - & - & 93.63±1.29 & 0.124±0.027 & 93.21±2.45 & 0.139±0.048 & - & - \\
Rater 2 & - & - & 92.90±1.90 & 0.145±0.035 & 92.35±2.50 & 0.169±0.055 & - & - \\
\bottomrule
\end{tabular}
\end{adjustbox}
\end{table}

When trained on D1 (Table~\ref{tab:train_d1}), MonoUNet outperformed all baselines across all metrics with Dice scores between 93.09\% and 94.68\%, and MASD values below 0.255 mm, comparable to inter-rater variability on D0 and D2.
TinyU-Net exhibited competitive performance on D0 (Dice: \(93.57\pm2.61\%\), MASD: \(0.147\pm0.046\) mm), but its performance significantly degraded on D2 (Dice: \(87.88\pm12.26\%\), MASD: \(0.190\pm0.190\) mm) and drastically collapsed on D3 (Dice: \(39.83\pm16.87\%\), MASD: \(1.218\pm1.244\) mm).

% \usepackage{booktabs}
% \usepackage{multirow}
% \usepackage[normalem]{ulem}
\begin{table}[]
\centering
\caption{Cross-dataset segmentation performance for models trained on D2 and evaluated on D0, D1, and D3. Results are reported as mean ± standard deviation for Dice and MASD (mm). Best results for each column are highlighted in bold and second-best are underlined. Inter-rater variability is measured against the retrospective manual segmentations. 
}
\label{tab:train_d2} 
\begin{adjustbox}{max width=\textwidth}
\begin{tabular}{@{}lcccccccc@{}}
\toprule
\multirow{2}{*}{\textbf{Config.}} & \multirow{2}{*}{\textbf{\begin{tabular}[c]{@{}c@{}}Params\\ (k)$\downarrow$\end{tabular}}} & \multirow{2}{*}{\textbf{\begin{tabular}[c]{@{}c@{}}FLOPS\\ (G)$\downarrow$\end{tabular}}} & \multicolumn{2}{c}{\textbf{D0}} & \multicolumn{2}{c}{\textbf{D1}} & \multicolumn{2}{c}{\textbf{D3}} \\ \cmidrule(l){4-9} 
 &  &  & \textbf{Dice}$\uparrow$ & \textbf{MASD}$\downarrow$ & \textbf{Dice}$\uparrow$ & \textbf{MASD}$\downarrow$ & \textbf{Dice}$\uparrow$ & \textbf{MASD}$\downarrow$ \\ \midrule
UNeXt~\cite{valanarasu_unext_2022} & 1471.94 & 4.59 & {\ul 94.32±3.06} & \textbf{0.116±0.062} & 52.50±22.76 & 3.523±4.523 & {\ul 90.38±4.90} & 0.244±0.202 \\
Med-NCA~\cite{kalkhof_med-nca_2023} & {\ul 70.02} & 310.31 & 91.37±8.26 & {\ul 0.133±0.254} & 5.64±9.56 & 18.452±7.849 & 89.96±6.40 & 0.267±0.238 \\
CMUNeXt-S~\cite{tang_cmunext_2023} & 417.50 & {\ul 2.18} & 94.09±2.89 & 0.134±0.037 & 64.57±22.07 & 4.373±5.251 & 91.57±4.84 & {\ul 0.258±0.203} \\
TinyU-Net~\cite{chen_tinyu-net_2024} & 481.17 & 3.33 & 93.63±3.11 & 0.145±0.042 & {\ul 80.92±10.43} & {\ul 0.332±0.941} & 83.77±10.40 & 0.299±0.194 \\
\textbf{MonoUNet} & \textbf{1.39} & \textbf{0.15} & \textbf{94.40±3.15} & 0.142±0.115 & \textbf{92.62±3.31} & \textbf{0.167±0.071} & \textbf{94.43±2.36} & \textbf{0.209±0.205} \\ \midrule
Rater 1 & - & - & 93.63±1.29 & 0.124±0.027 & 93.67±1.92 & 0.133±0.034 & - & - \\
Rater 2 & - & - & 92.90±1.90 & 0.145±0.035 & 91.37±3.42 & 0.192±0.067 & - & - \\
\bottomrule
\end{tabular}
\end{adjustbox}
\end{table}

\begin{figure}[h]
    \centering
    \includegraphics[width=0.97\linewidth]{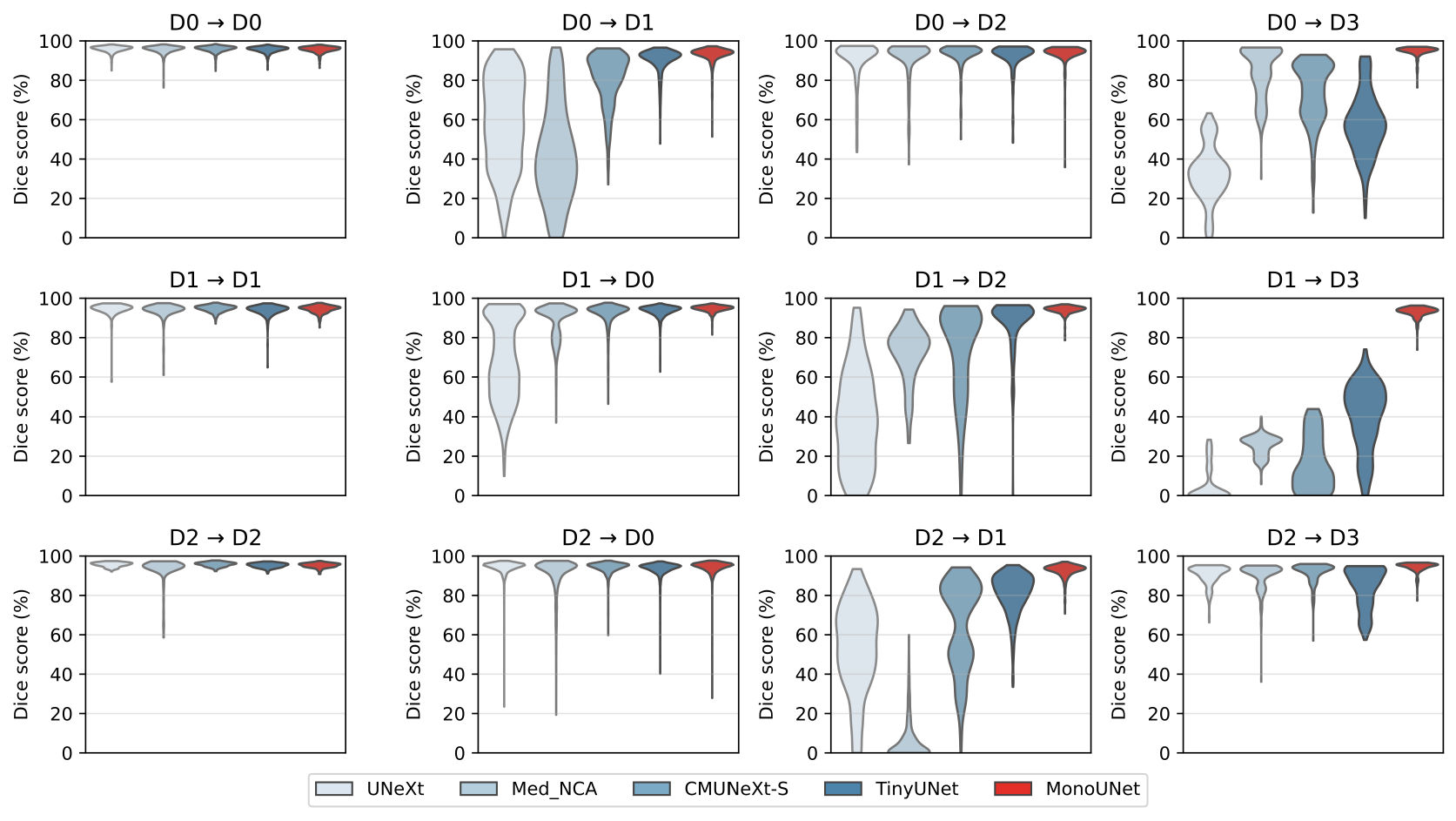}
    \caption{Cross-dataset Dice score distributions for the three training scenarios. \textbf{Top row} shows models trained on D0, \textbf{middle row} shows models trained on D1, and \textbf{bottom row} shows models trained on D2 and tested on D0, D1, and D3.}
    \label{fig:dice_distribution}
\end{figure}

When trained on D2, the lowest quality dataset (Table~\ref{tab:train_d2}), MonoUNet achieved Dice scores between 92.62\% and 95.23\% with corresponding MASD values below 0.21 mm, still comparable to inter-rater variability on D0 and D1.
MonoUNet outperformed all baselines across all metrics except MASD on D0 where UNeXt and CMUNeXt-S achieved lower values (\(0.116\pm0.062\) mm and \(0.133\pm0.254\) mm vs \(0.142\pm0.115\) mm).

Fig.~\ref{fig:dice_distribution} shows the distribution of the Dice scores across the three training scenarios. 
MonoUNet consistently exhibited stable high median Dice scores with narrow distributions across all train–test combinations.
In contrast, all baselines had variable performance distributions across different scenarios despite achieving very high performance on the validation sets (i.e., first column of Fig.~\ref{fig:dice_distribution}).
% The worst performance drop was observed for training scenario \(D1\rightarrow D3\) where all the baseline models achieved Dice values below 80\% while MonoUNet maintained its performance (Dice > 90\%).

\subsection{Ablation study}
Table~\ref{tab:ablation} shows the contribution of individual modules within MonoUNet.
The baseline U-Net architecture achieved a Dice score of 93.84\% and MASD of 0.136 mm.
Halving the decoder size has no effect on model performance while reducing model size by close to 40\% (from 46 to 27 million parameters).
However, reducing the model size further by aggressively reducing the number of channels (MonoUNetBase) degrades performance to a Dice value of 89.31\% and MASD of 0.218 mm.
Introducing the Mono block at the first encoder stage with a single learnable scale (MonoUNetE1) substantially improves performance to 92.07\% Dice and 0.167 MASD.
However, injecting single-scale features into multiple encoder stages (MonoUNetE123) results in reduces performance relative to MonoUNetE1.
Introducing multi-scale features (MonoUNetE123V2) restores and improves performance yielding a Dice score of 92.60\% and a MASD of 0.177 mm.
Incorporating the gating mechanism (MonoUNetE123V2Gated) further improves performance slightly to 92.68\% Dice and 0.175 mm MASD.
Finally, adding scaling and rotation data augmentation (MonoUNetE123V2GatedDA) further improves performance to a Dice score of 94.14\% and an MASD of 0.149 mm.
Overall, the ablation results indicate that both multi-scale feature integration and gated feature fusion contribute to improved segmentation accuracy.

% Please ad% Please add the following required packages to your document preamble:
% \usepackage{booktabs}
\begin{table}[h]
\centering
\caption{Ablation study of MonoUNet individual modules. Results are reported as mean ± standard deviation for Dice and MASD (mm). Models were trained on D1 and tested on D2. The proposed method is highlighted in \textbf{bold}.}
\label{tab:ablation}
\small\selectfont
\begin{adjustbox}{max width=\textwidth}
\begin{tabular}{@{}lcccc@{}}
\toprule
\textbf{Config.} & \textbf{\begin{tabular}[c]{@{}c@{}}Params\\ (k)$\downarrow$\end{tabular}} & \textbf{\begin{tabular}[c]{@{}c@{}}FLOPS\\ (G)$\downarrow$\end{tabular}} & \textbf{Dice}$\uparrow$ & \textbf{MASD}$\downarrow$ \\ \midrule
U-Net & 46319 & 160.55 & 93.84±4.26 & 0.136±0.037 \\
U-Net (half decoder) & 27961 & 33.45 & 93.88±4.02 & 0.138±0.044 \\
MonoUNetBase & 1.14 & 0.04 & 89.31±6.80 & 0.218±0.086 \\ \midrule
MonoUNetE1 & 1.21 & 0.04 & 92.07±4.74 & 0.167±0.060 \\
MonoUNetE123 & 1.21 & 0.04 & 91.21±5.73 & 0.181±0.067 \\
MonoUNetE123V2 & 1.30 & 0.04 & 92.60±2.72 & 0.177±0.058 \\
MonoUNetE123V2Gated & 1.39 & 0.05 & 92.68±2.84 & 0.175±0.060 \\
\textbf{MonoUNetE123V2GatedDA (MonoUNet)} & 1.39 & 0.05 & \textbf{94.14±2.05} & \textbf{0.149±0.038} \\ \bottomrule
\end{tabular}
\end{adjustbox}
\end{table}

\subsection{Clinical utility of the automated segmentations}

\begin{figure}[h]
    \centering
    \includegraphics[width=0.9\linewidth]{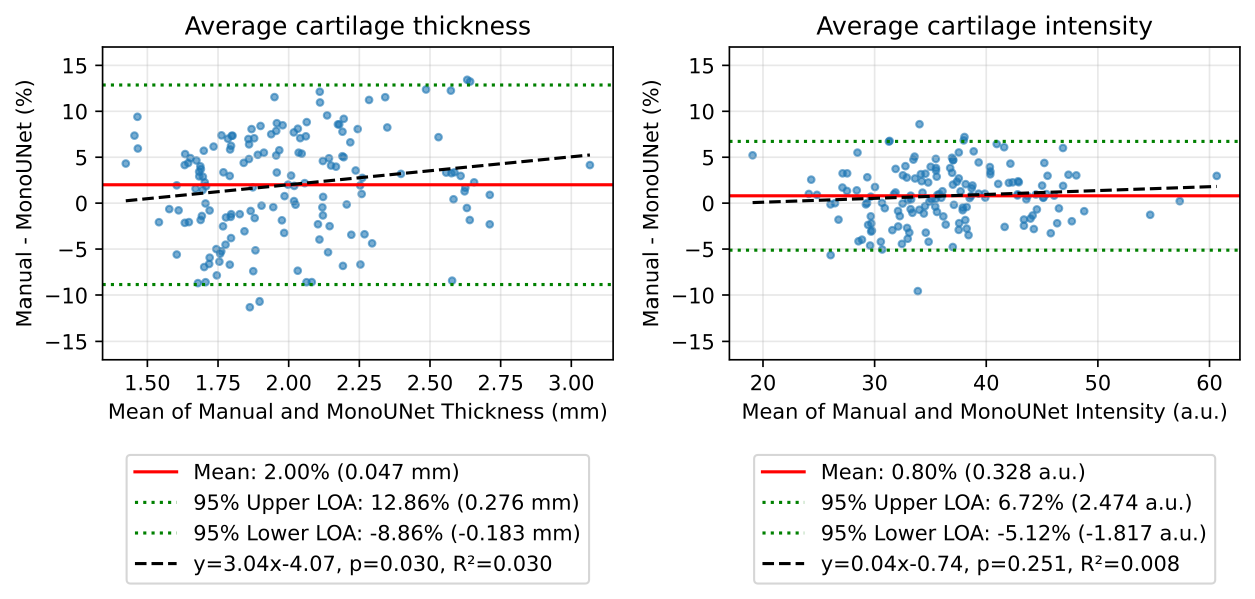}
    \caption{Bland-Altman plots showing the agreement between cartilage thickness and intensity measured from manual and automated segmentations.}
    \label{fig:bland_altman}
\end{figure}

Fig.~\ref{fig:bland_altman} shows Bland-Altman plots comparing the average cartilage thickness and intensity outcomes of both the manual and MonoUNet segmentations.
For average cartilage thickness, there was excellent reliability between the manual and MonoUNet segmentations with ICC\(_{2,k}\) = 0.96 (95\% CI [0.93, 0.97]), mean bias of 2.00\% (95\% LOA [\(-8.86\%\), \(12.86\%\)]), and a statistically significant but weak proportional bias (\(R^2 = 0.030\), \(p = 0.030\)).
For average echo intensity, there was excellent reliability between the manual and MonoUNet segmentations with ICC\(_{2,k}\) = 0.99 (95\% CI [0.99, 0.99]), mean bias of 0.80\% (95\% LOA [\(-5.12\%\), \(6.72\%\)]), and no significant proportional bias (\(R^2 = 0.008\), \(p = 0.251\)).

\subsection{Qualitative results}

\begin{figure}[]
    \centering
    \includegraphics[width=\linewidth]{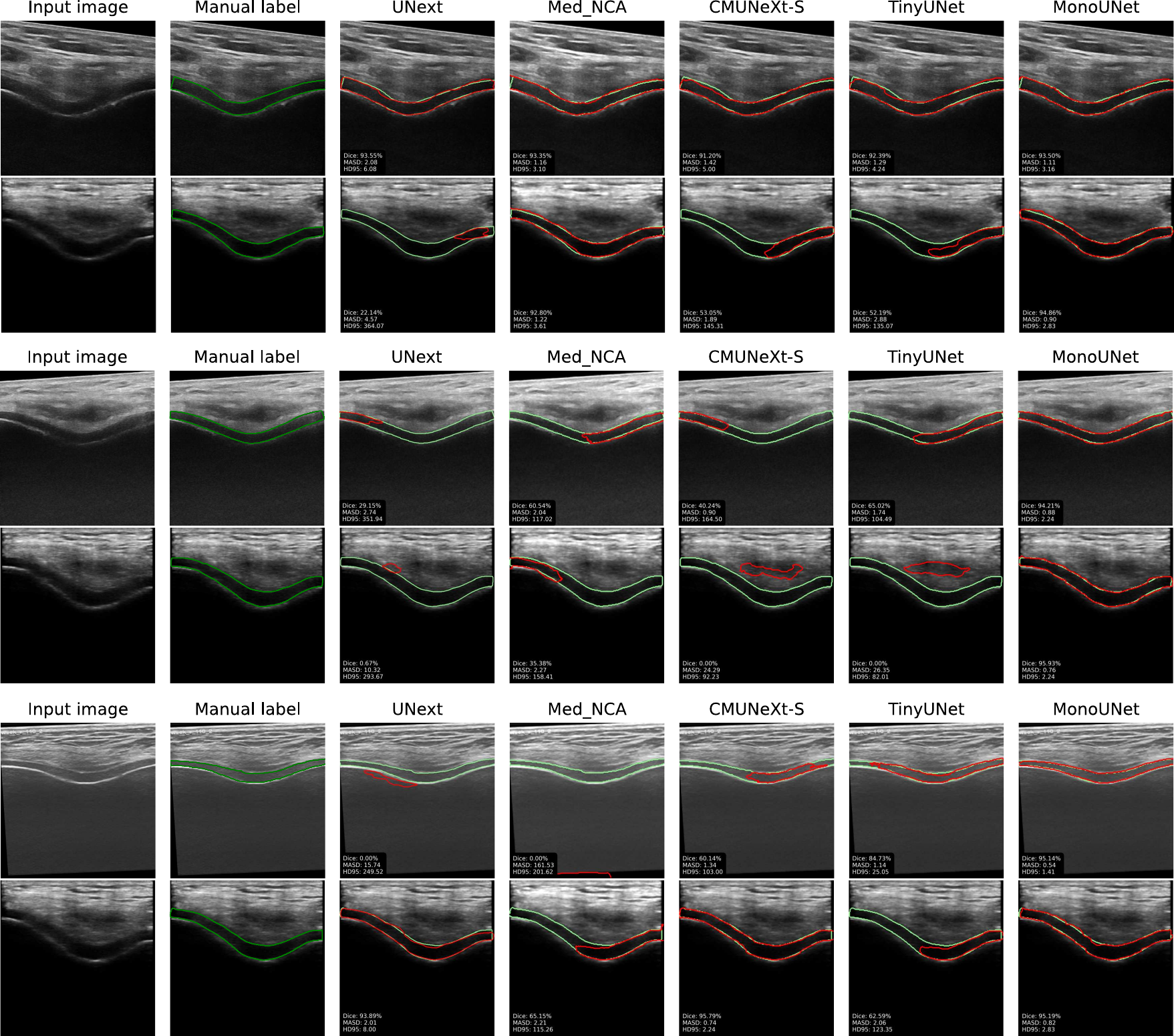}
    \caption{Representative qualitative results on unseen test images.
    Manual annotations are shown in green and model predictions in red.
    Rows 1–2 show models trained on D0 and tested on D2 and D3; rows 3–4 show models trained on D1 and tested on D2 and D3; rows 5–6 show models trained on D2 and tested on D1 and D3.
    }
    \label{fig:qualitative}
\end{figure}

Fig.~\ref{fig:qualitative} shows representative qualitative segmentation results on unseen test images for the three training scenarios.
MonoUNet segmentations closely followed the manual labels, even under challenging image quality and acquisition conditions such as the middle rows (\(D1\rightarrow D2\) and \(D1\rightarrow D3\)).
In contrast, several baseline architectures exhibited partial segmentations, boundary leakage, or fragmented predictions.
These qualitative observations were consistent with the quantitative performance trends and the Dice score distributions shown in Fig.~\ref{fig:dice_distribution}.

\section{Discussion}
Existing lightweight models exhibit limited cross-device generalization for knee cartilage segmentation as shown in Fig.~\ref{fig:dice_distribution}.
Their performance tends to degrade less when the test data closely matches the training distribution.
This is evident in scenarios such as \(D0\rightarrow D2\) and \(D2 \rightarrow D0\), where the datasets share the same subjects, as well as \(D2\rightarrow D3\), where the image quality is relatively similar.
However, performance degrades substantially under more pronounced domain shifts such as \(D1\rightarrow D3\).
MonoUNet, on the other hand, maintains consistently high segmentation accuracy and robustness in all training scenarios, generally outperforming existing lightweight architectures despite having orders of magnitude fewer parameters.
The results demonstrate that MonoUNet achieves a favorable balance between model compactness, segmentation accuracy, and robustness to domain shift.

We attribute MonoUNet's performance to the integration of trainable local phase features, which emphasize structural information that is inherently less sensitive to absolute intensity and contrast variations (Fig.~\ref{fig:qualitative}).
This property is well aligned with the physics of ultrasound imaging, where speckle, gain, and device-specific processing can significantly alter image appearance without changing underlying anatomy.
The ablation study (Table~\ref{tab:ablation}) further supports this interpretation, showing consistent performance gains when phase features and gated fusion are incorporated.

These results are consistent with existing literature exploiting local phase information in ultrasound image analysis~\cite{hacihaliloglu_2a-4_2006,hacihaliloglu_bone_2009} including knee cartilage segmentation~\cite{desai_knee-cartilage_2019,harkey2022validating}.
However, these approaches manually tune the local phase filters which is challenging, requires significant expertise, and is subjective.
\citet{moya-sanchez_trainable_2021} proposed the first work exploring learning monogenic filters on natural image classification tasks.
However, their approach learns only a single scale monogenic filter placed right before the input of the deep learning network.
In contrast, for MonoUNet, we learn multi-scale local phase features and inject them into multiple encoder stages to regulate the base encoder features rather than forcing the model to rely only on the local phase features.
This allows flexibility in the model to selectively exploit the features and we found it to perform better (Table~\ref{tab:ablation}).

Nevertheless, future work will focus on reducing variability in cartilage thickness measurements to improve agreement.
In addition, other phase-based features, such as phase symmetry, asymmetry, and congruency, alongside the local phase features will be investigated.
Furthermore, large-scale validation on diverse populations and devices is still required to fully assess the generalizability of MonoUNet.

\section{Conclusion}

We presented MonoUNet, an ultra-compact neural network for automated knee cartilage segmentation in POCUS devices.
By integrating trainable local phase features into a minimal U-Net backbone, MonoUNet achieves robust segmentation performance across multiple devices and acquisition conditions while maintaining real-time inference on handheld hardware.
These results highlight the potential of explicitly encoding structural priors to overcome domain shift in resource-constrained medical imaging applications and support the broader adoption of ultrasound-based knee osteoarthritis assessment.

\section{Acknowledgments}
This work was supported by the Canadian Foundation for Innovation-John R. Evans Leaders Fund (CFI-JELF) program [Grant ID 42816, AWD-023869 CFI].
We acknowledge the support of the Natural Sciences and Engineering Research Council of Canada (NSERC), [RGPIN-2023-03575 AWD-024385]. Cette recherche a été financée par le Conseil de recherches en sciences naturelles et en génie du Canada (CRSNG), [RGPIN-2023-03575, AWD-024385].
We acknowledge the support provided by the Canadian Consortium of Clinical Trial Training (CANTRAIN) platform, Michael Smith Health Research British Columbia, and the Minimally Invasive Image Guided Procedure Lab (MIIPs), at the University of British Columbia School of Biomedical Engineering.

\section{Conflict of Interest Statement}
The authors have no competing interests to declare that are relevant to the content of this article.

\section{Data Availability Statement} The dataset used in this paper will be available upon reasonable request.
The code is publicly available at \href{https://github.com/alvinkimbowa/monounet}{https://github.com/alvinkimbowa/monounet}.

\section{Human and Animal Rights}
Ethics approval was obtained to collect data and informed consent was obtained from all individual participants included in the study.

% %% The Appendices part is started with the command \appendix;
% %% appendix sections are then done as normal sections
% \appendix
% \section{Appendices}
% \subsection{Deployment Setup and Details}

% Before deployment, we used PyTorch to jit trace to optimize the models.
% Note that no further optimization (quantization, pruning) were done.
% We then integrated the models into the Clarius Cast Android mobile application\footnotemark.
% \footnotetext{https://github.com/clariusdev/cast.git}
% We deployed the application on an 4GB RAM Samsung Galaxy Tab A9+.
% Inference time was measured as the time taken for the model to make a prediction given an input image.
% Pre-processing time included time taken to crop, resize, and normalize the input image.
% Post-processing time measured the time taken to extract the largest component from the prediction and overlay the prediction onto the image.


\bibliographystyle{elsarticle-num-names} 
\bibliography{refs}

\end{document}